\documentclass[aps,prl,amsmath,amssymb,twocolumn]{revtex4-1}

\usepackage{graphicx}
\usepackage{dcolumn}

\usepackage{amsfonts}
\usepackage{bm}

\newcommand{\ot}{{\,\otimes\,}}
\newcommand{{\Cd}}{{\mathbb{C}^d}}

\def\oper{{\mathchoice{\rm 1\mskip-4mu l}{\rm 1\mskip-4mu l}%
{\rm 1\mskip-4.5mu l}{\rm 1\mskip-5mu l}}}
\def\<{\langle}
\def\>{\rangle}

\begin{document}


\title{Non-Markovianity as a Resource for Quantum Technologies}

\author{Bogna Bylicka}
\author{ Dariusz Chru\'sci\'{n}ski}
\affiliation{Institute of Physics, Faculty of Physics, Astronomy and Informatics, Nicolaus Copernicus University, Grudziadzka 5, 87-100 Toru\'{n}, Poland}

\author{Sabrina Maniscalco}
\email{s.maniscalco@hw.ac.uk}
\affiliation{SUPA, EPS/Physics, Heriot-Watt University, Edinburgh, EH14 4AS, United Kingdom}
\affiliation{Department of Physics and Astronomy, University of Turku, 20014
Turku, Finland}

\date{\today}

\begin{abstract}
 Quantum technologies rely on the ability to coherently manipulate, process and transfer information, encoded in quantum states, along quantum channels. Decoherence induced by the environment introduces errors, thus setting limits on the efficiency of any quantum-enhanced protocol or device. A fundamental bound on the ability of a noisy quantum channel to transmit quantum (classical) information is given by its quantum (classical) capacity \cite{Nielsen}. Generally, the longer is a quantum channel the more are the introduced errors, and hence the worse is its capacity. In this Letter we show that for non-Markovian quantum channels \cite{NMQJ, NMBLP, NMCV, NMWolf, NMRHP, NMFisher, NMLuo, NMCK, NMNatPhys} this is not always true: surprisingly the capacity of a longer channel can be greater than the one of a shorter channel. We introduce a general theoretical framework linking non-Markovianity to the capacities of quantum channels, and demonstrate in full generality how harnessing non-Markovianity may improve the efficiency of quantum information processing and communication.
\end{abstract}
\maketitle

The description of quantum systems interacting with their environment is the central objective of the theory of open quantum systems \cite{BPBook}. During the last few years there has been an increasing interest in open quantum systems with memory, also known as non-Markovian open quantum system, due to both fundamental and applicative reasons. From a fundamental point of view the study of quantum systems interacting with structured environments, while presenting considerable difficulties from a theoretical point of view, is of crucial importance for the realistic description of a variety of physical systems such as photonic band gap materials, quantum biological systems and complex quantum networks, solid state systems (e.g., SQUIDs and Josephson junctions), and ultracold gases. From an applicative point of view, the increasing ability in reservoir engineering techniques paves the way to new methods of decoherence control based on the manipulation and modification of properties of the environment such as its frequency spectrum \cite{REItano, REIons, REVerstraete}.

Non-Markovianity is a multifaceted and complex phenomenon which cannot be simply grasped by looking at specific instances and cannot be generally traced back to a single unique feature of the environment. Several measures of non-Markovianity have been introduced, based on distinguishability of quantum states as measured by trace distance \cite{NMBLP} or fidelity \cite{NMCV}, semigroup property \cite{NMWolf} or divisibility \cite{NMRHP} of the dynamical map, Fisher information \cite{NMFisher}, or quantum mutual information \cite{NMLuo}. In general these measures or witnesses do not coincide and examples of differences between them have been reported even for simple open quantum systems \cite{PRAPinja}. This is an obvious consequence of the fact that reservoir memory may have different effects  on different dynamical properties which one may want to harness for certain specific purposes. Rather than being a problematic aspect, we believe that this richness and variety constitutes the power of non-Markovian open quantum systems.

Very recently the role of structured environments and non-Markovianity in quantum metrology \cite{NMRMetrology}, quantum key distribution \cite{NMRQKD}, quantum teleportation \cite{NMRTeleportation}, entanglement generation \cite{NMREG}, optimal control \cite{NMOC}, and quantum biology \cite{NMresourseQbio, NMHuelgaQbio}, has started to be investigated, showing  with increasing evidence that non-Markovian quantum channels may be advantageous compared to Markovian ones. However, to date there is no general theory linking non-Markovian dynamics with an increase in the efficiency of quantum information processing and communication. This is exactly the scope of this Letter. Our main result is the identification of specific features of non-Markovianity which lead to an increase in the capacities of quantum channels compared to the corresponding Markovian ones.

The general scenario typical of quantum information processing and communication sees Alice and Bob at the opposite ends of a quantum channel, the former sending information (classical or quantum) and the latter one receiving it. The maximum amount of information that can be reliably transmitted along a noisy quantum channel is known as the channel capacity. In this Letter we will be concerned with two types of capacities, the entanglement assisted capacity $C_{ea}$ and the quantum capacity $Q$ \cite{QChHolevo} . The first quantity sets a bound on the amount of classical information which can be transmitted along a quantum channel when one allows Alice and Bob to share an unlimited amount of entanglement. It  is defined in terms of the quantum mutual information $I (\rho, \Phi_t)$ between the input and the output of the channel
\begin{equation}
C_{ea}(\Phi_t)=\sup_{\rho} I(\rho, \Phi_t), \label{eq:Cea}
\end{equation}
where
\begin{equation}\label{}
    I(\rho, \Phi_t)=S(\rho)+S(\Phi_t \rho )-S(\rho,\Phi_t),
\end{equation}
with $S(\rho)$ the von Neumann entropy of the input state, $S(\Phi_t \rho)$ the entropy of the output state and $S(\rho,\Phi_t)$ the entropy exchange. When referring to a quantum channel $\Phi_t$ we explicitly indicate the time $t$ which tells us for how long the quantum state, encoding classical or quantum information, is subjected to environmental noise. In experimental implementations of quantum protocols, e.g. with trapped ion systems, $t$ is the duration of the experiment and it is obviously connected to the length of the channel, i.e. the length of an optical fiber in optical systems. The quantum capacity $Q$, on the other hand, gives the limit to the rate at which quantum information can be reliably sent down a quantum channel. For single use of the channel, it is defined in terms of the coherent information $I_c (\rho, \Phi_t)$ as follows \cite{QCLoyd}
\begin{equation}\label{eq:Q}
    Q(\Phi_t)=\sup_{\rho} I_c(\rho, \Phi_t),
\end{equation}
with
\begin{equation}\label{}
    I_c(\rho, \Phi_t)=S(\Phi_t \rho)-S(\rho,\Phi_t).
\end{equation}
More in general, for $n$ successive uses of the channel, the quantum channel capacity is defined as $Q(\Phi_t) = \lim_{n \rightarrow \infty} [\max_{\rho_n }I_c(\rho_n,\Phi_t ^{\otimes n})]/n$. We note that, contrarily to the entanglement assisted capacity, the quantum channel capacity is in general not additive. However, for degradable channels \cite{QCDegradable}, the general definition coincide with Eq. (\ref{eq:Q}), and additivity holds.

One of the central results of quantum information theory is the quantum data processing inequality \cite{QDP_SchuNiel} which, intuitively, says that processing quantum information reduces the amount of correlations between input and output. More precisely, given the quantum channels ${\cal E}_1$, ${\cal E}_{12}$, and their concatenation ${\cal E}_2 = {\cal E}_{12} {\cal E}_1$,  we have
\begin{equation}\label{}
    I_c(\rho, {\cal E}_2) \le I_c(\rho, {\cal E}_1).
\end{equation}
For divisible channels $\Phi_{t} = \Phi_{t,s} \Phi_s $, with $s\le t$, the data processing inequality implies $ I_c(\rho, \Phi_t) \le I_c(\rho, \Phi_s)$. A similar inequality holds for the mutual information, i.e. $ I(\rho, \Phi_t) \le I(\rho, \Phi_s) $. As a consequence, for divisible quantum channels, both the entanglement assisted capacity $C_{ea}(\Phi_t)$ and the quantum capacity $Q (\Phi_t)$ decrease monotonically with time. Divisibility is, however, a property of Markovian dynamical maps. All existing measures of non-Markovianity \cite{NMBLP, NMCV, NMWolf, NMRHP, NMFisher, NMLuo} are based on non-monotonic behaviour of certain quantities which occurs when the divisibility property is violated. Following the same line of reasoning, we define here two  new measures of non-Markovianity based on the non-monotonic behavior of the quantum and entanglement assisted capacities,
\begin{equation}\label{NQ}
\mathcal{N}_Q=\int_{\frac{d Q(\Phi_t)}{dt} >0} \frac{d Q(\Phi_t)}{dt} dt,
\end{equation}
and
\begin{equation}\label{NC}
\mathcal{N}_{C}=\int_{ \frac{d C_{ea}}{dt}(\Phi_t) >0} \frac{dC_{ea}(\Phi_t)}{dt} dt,
\end{equation}
where the integrals above are extended to all time intervals over which $dQ/dt$  and $dC_{ea}/dt$ are positive.

These two measures of non-Markovianity, in general, do not coincide even for degradable channels and in fact they distinguish between two different types of resources, being related to revivals of correlations which can be used to transfer either classical information or quantum information down a quantum channel. The distinction between these quantities is actually quite subtle. We notice indeed that as $ I(\rho,\Phi_t) = S(\rho)+I_c(\rho, \Phi_t)$, with $\rho$ the input state, we have $\frac{d}{dt} I(\rho,\Phi_t) = \frac{d}{dt} I_c(\rho, \Phi_t)$. Therefore a measure based, e.g., on the violation of the data processing inequality for certain non divisible maps would not be able to distinguish between an increase in the different types of correlations. The optimizing state in the definitions (\ref{eq:Cea})-(\ref{eq:Q}), however, is time dependent and does not coincide for the two quantities, hence  $dQ/dt \neq dC_{ea}/dt$.

The two measures introduced above are very useful to illustrate how specific features of non-Markovianity may be seen as a resource for quantum information processing and communication. More precisely, Markovian dynamical maps characterized by constant decoherence rates generally lead to irretrievable deterioration of the channel capacity as the length of the channel increases. On the contrary, when compared to their corresponding Markovian maps, non-Markovian dynamical maps charactertised by time-dependent decoherence rates may lead to (i)  increase of the channel capacities for a given channel length, (ii) revivals of the channel capacities, hence increasing the values of channel lengths over which the capacities are non zero, (iii)  length-independent finite-capacity channels (residual channel capacity), i.e., channels for which the quantum and/or entanglement assisted capacity remains unchanged and positive, after a certain threshold length. In the following we will illustrate these points by looking at three exemplary types of exact, and therefore non-Markovian, quantum channels: the dephasing channel, the amplitude damping channel in a Lorentzian environment and the amplitude damping channel in a photonic band gap.

Let us begin by considering the dephasing channel for a qubit, described by the following formula
\begin{equation}\label{DEF}
    \Phi^D_t(\rho) =\left(
                     \begin{array}{cc}
                       \rho_{11} & \rho_{12} e^{-\Gamma(t)} \\
                        \rho_{21} e^{-\Gamma(t)}& \rho_{22} \\
                     \end{array}
                   \right),
\end{equation}
where $\Gamma(t)=\int_0^t \gamma(\tau) d\tau$, $\gamma(t)$ is a time-dependent decoherence rate, and $\rho_{ij}$ are the density matrix elements of the initial state $\rho$. The formula (\ref{DEF}) provides a legitimate quantum channel iff $\Gamma(t)\geq 0$. Note, that decoherence rate $\gamma(t)$ needs not be positive. If $\gamma(t) \geq 0$, then the channel is divisible.
\begin{figure}
\label{fig1}
\includegraphics[width=0.45\textwidth]{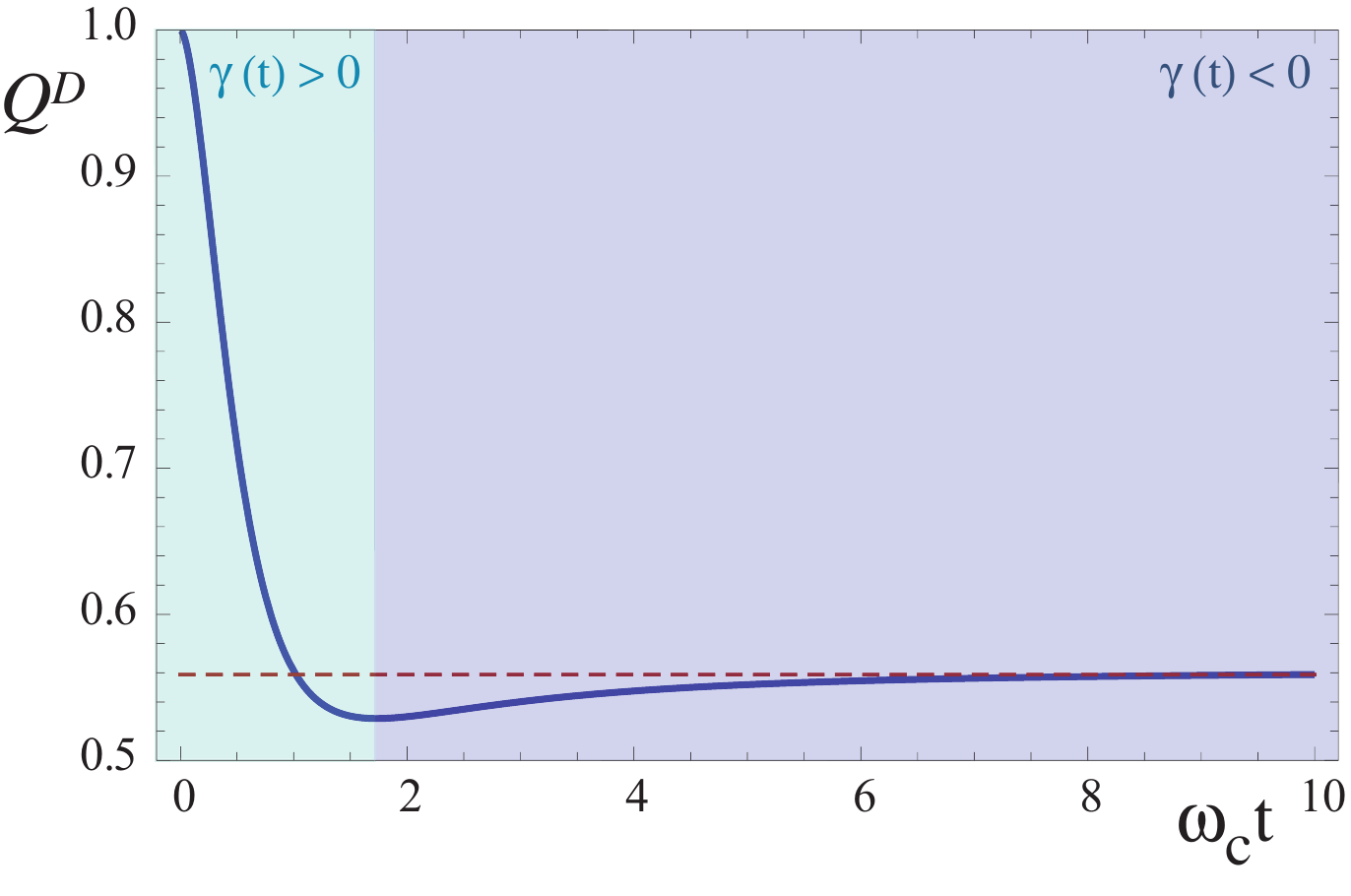}
\caption{
{\bf Quantum channel capacity $Q^D$ as a function of time or, equivalently, of the channel length for the exact dephasing model}. We consider a super-Ohmic reservoir spectrum with $s=3$ and $\gamma_M/ \omega_c = 0.1$. The light-blue shaded region corresponds to positive values of the dephasing rate $\gamma(t)$, for which $d Q^D/dt < 0$ while the dark-blue shaded region corresponds to an interval of time for which $\gamma(t) < 0$ and $d Q^D/dt \ge 0$.}
\end{figure}
The dephasing channel is degradable for all admissible $\gamma(t)$, i.e. whenever $\Gamma(t) \ge 0$, which simplifies the calculations of the quantum capacity. In this case, indeed, we find that the state optimizing the coherent information in the definition of the quantum capacity does not depend either on time or on the specific properties of the environmental spectrum.
Having this in mind one can show that $Q$ takes a simple analytical formula  \cite{QCDegradable} and that
the measure $\mathcal{N}_Q$ has nonzero value if and only if $\gamma(t)<0$, i.e., whenever the dynamical map is non-divisible. Moreover, a similar calculation for the entanglement assisted capacity shows that
$C_{ea}^D(t) = 1+ Q^D(t)$. It follows immediately that $\mathcal{N}_Q=\mathcal{N}_{C}$. For dephasing channels all known measures of non-Markovianity vanish if and only if the channel is divisible. In the Supplementary Information we discuss in detail the comparison between $\mathcal{N}_Q$, $\mathcal{N}_{C}$ and all other non-Markovianity measures for both the dephasing and the amplitude damping channel.

A very interesting features of $\mathcal{N}_Q$ and $\mathcal{N}_{C}$ is that, due to the additivity of $Q$ and $C_{ea}$, they satisfy the property $\mathcal{N}_Q ( \Phi_t ^{\otimes n}) = n  \mathcal{N}_Q ( \Phi_t )$, which allows to calculate straightforwardly the non-Markovianity measure of $n$ qubits dephasing in identical uncorrelated environments from the non-Markovianity measure of a single dephasing qubit.

In Fig. 1 we show the behavior of the quantum channel capacity $Q^D$ as a function of time or, equivalently, of the channel length for the exact model of dephasing with Ohmic reservoir spectrum of the form $J(\omega)= \gamma_M (\omega/\omega_c)^{s} e^{-\omega/\omega_c }$, with $\omega_c$ the cutoff frequency, $\gamma_M$ the coupling constant, and $s$ the Ohmicity parameter. For zero temperature bosonic environments $\Phi^D_t$ is divisible if and only if $0< s \le 2 $ \cite{NMSpectrum}. We can change the non-Markovian character of the channel by changing the Ohmicity paramter $s$. This can be experimentally implemented, e.g., with impurities in ultracold atomic gases as demonstrated in Ref. \cite{NMBEC}.
 Figure 1 illustrates two important features of the non-Markovian character of the quantum channel, namely, the non-monotonicity of $Q^D$, which initially decreases with time but then starts increasing again after a certain threshold value of time or channel length, and the existence of residual quantum channel capacity. This should be contrasted with the Markovian dephasing channel for which $Q^D$ exponentially decay to zero with positive constant decay rate $\gamma_M$.

\begin{figure}
\includegraphics[width=0.45\textwidth]{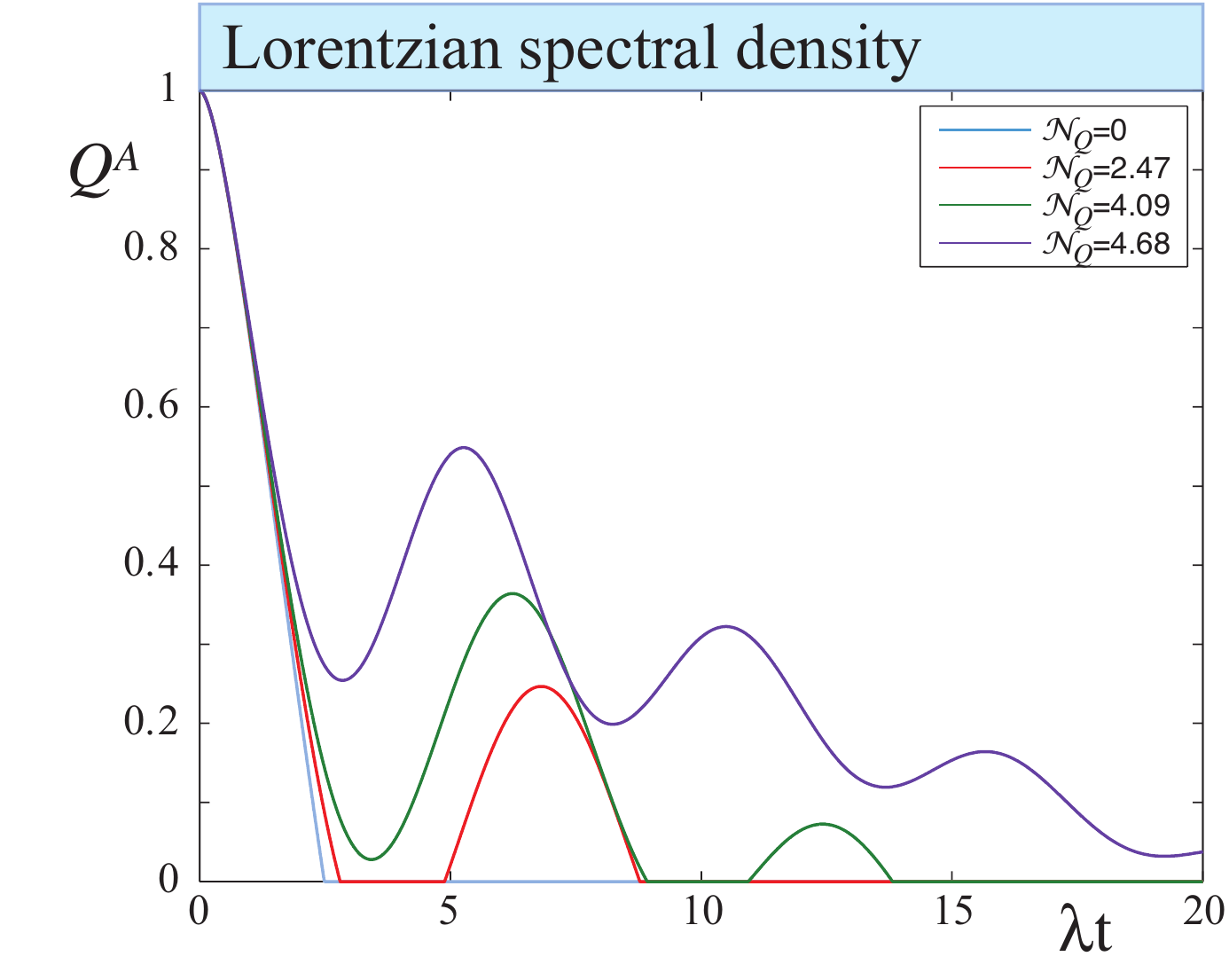}
\caption{
{\bf Quantum channel capacity $Q^A$ as a function of time or, equivalently, of the channel length for the exact amplitude damping model with Lorentzian reservoir spectrum}. In the plot we have set $\lambda/\gamma_M=0.06$, with detuning parameters $(\omega_0-\omega_c)/\lambda$= 3 (blue line), 5 (red line), 6 (green line), and 8 (violet line), where $\omega_0$ is the Bohr frequency of the two-level system. Increasing values of detuning parameters correspond to increasing values of the non-Markovianity measure ${\cal N}_Q$ (see inset), and correspondingly to higher and higher values of the quantum channel capacity $Q^A$.}
\end{figure}

The second example we consider is the amplitude damping channel described by
\begin{equation}
    \Phi^A_t(\rho)=\left(
              \begin{array}{cc}
                1-|G(t)|^2 \rho_{22} & G(t) \rho_{12} \\
                G^*(t) \rho_{12}^* & |G(t)|^2 \rho_{22} \\
              \end{array}
            \right),
\end{equation}
where the complex function $G(t)$ satisfies $|G(t)|\leq 1$ for all $t \geq 0$, and it is directly related to the reservoir spectral density $J(\omega)$ (See Supplementary Information).


The amplitude damping channel is  degradable for $|G(t)|^2>\frac{1}{2}$, while for $|G(t)|^2 \le \frac{1}{2}$ is non-degradable with zero quantum capacity. Quantum and entanglement assisted capacities in this case are calculated numerically \cite{Cap_GioFaz, Cap_BAF}. The states optimizing $I_c(\rho, \Phi_t)$ and $I(\rho, \Phi_t)$ are now time-dependent, but the optimization problem is still solvable.

In Fig. 2 we plot the quantum channel capacity $Q^{A}$ for the amplitude damping channel with Lorentzian reservoir spectral density of the form $J(\omega)=\gamma_M \lambda^2/2 \pi [(\omega-\omega_c)^2+\lambda^2]$, typical of cavity quantum electrodynamics. Interesting features emerging from this figure are the increase of $Q^A$ for increasing values of the non-Markovianity measure ${\cal N}_Q$, and the appearance of revivals of $Q^A$ after intervals of time/channel length for which $Q^A =0$ (See Fig. 2, red and green curves). Once more this should be contrasted with the Markovian behavior of the quantum channel capacity. The corresponding Markovian dynamics is given by $|G_M (t)|^2 = \exp (- \gamma_M t)$, hence for the value of parameters of Fig. 2, $Q^A=0$ for $\lambda t \ge 0.04$.

In Fig. 3 we plot again the quantum channel capacity $Q^{A}$, focussing now on the photonic band gap model of Ref. \cite{PBG}. This model is an intrinsically non-Markovian one and it does not admit a Markovian limit. This figure illustrates that the existence of length-independent finite-capacity channels, is not an exclusive feature of non-Markovian dephasing channels. Indeed, while such a situation never occurs for the amplitude damping channel with Lorenztian spectral density (See Fig. 2), it is found again in the photonic band gap model, where it stems from the well known phenomenon of population trapping.

For amplitude damping channels the non-Markovianity measures associated to $Q$ and $C_{ea}$, defined in Eqs. (\ref{NQ})-(\ref{NC}), are not simply related. In general $Q^A \le C^A_{ea}$, and the entanglement assisted capacity shows similar features to the ones that we have seen for the quantum capacity. Finally, we note that, for the amplitude damping channel, there exist values of parameters for which the non-Markovianity measures of Refs \cite{NMBLP, NMRHP} are non zero, but still ${\cal N}_Q =0  $, proving that these measures in general are not equivalent (See Supplementary Information).

\begin{figure}
\includegraphics[width=0.45\textwidth]{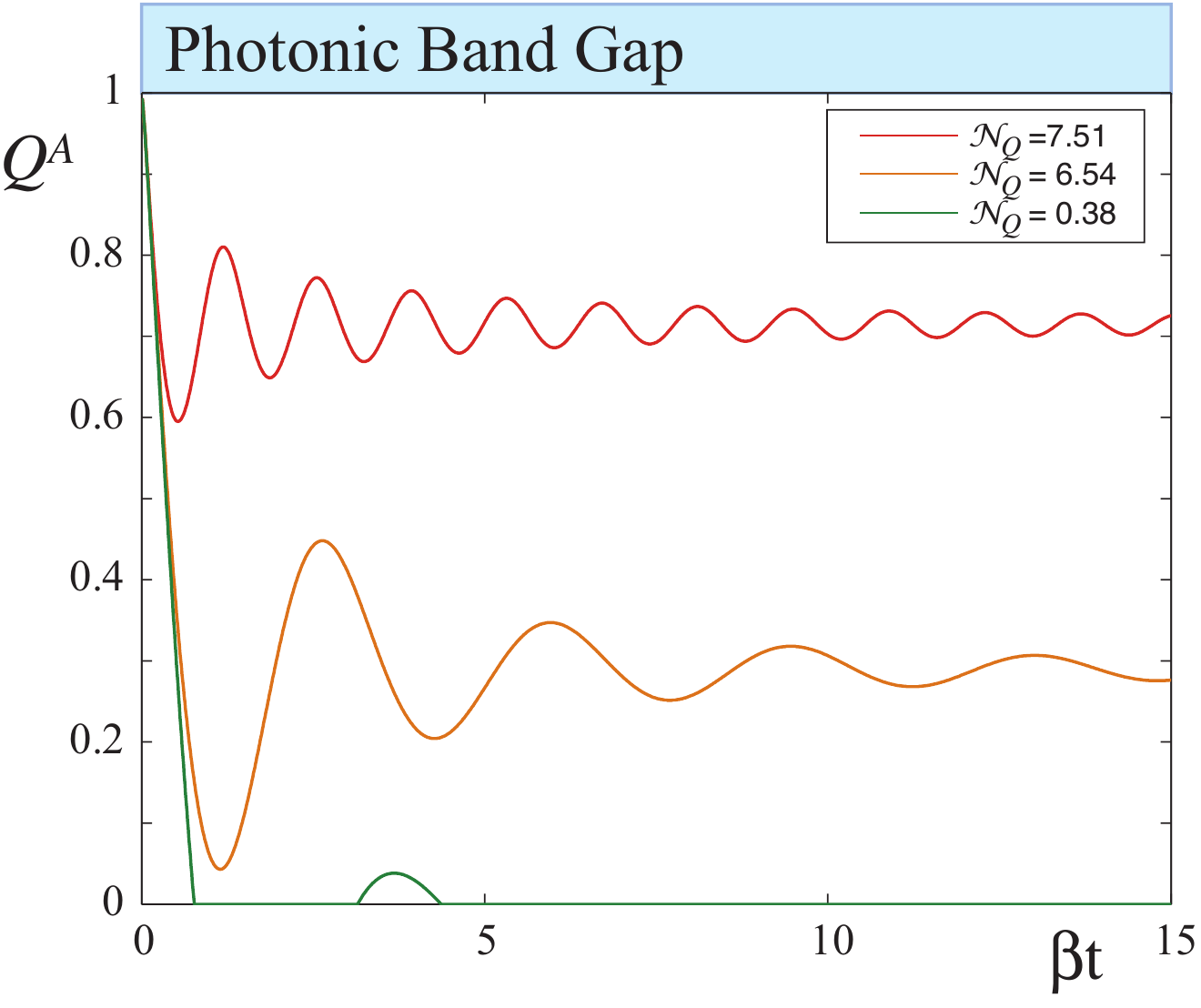}
\caption{
{\bf Quantum channel capacity $Q^A$ as a function of time or, equivalently, of the channel length for the exact amplitude damping model in a photonic band gap}. The detuning parameters are $(\omega_0-\omega_e)/ \beta = -4$ (red line), -1 (yellow line), and 0 (green line), where $\omega_0$ is the Bohr frequency of the two-level system and $\omega_e$ is the edge frequency of the band gap.}
\end{figure}

In the quest for realistic large scale implementations of quantum devices for quantum technologies one of the major existing challenges is the identification of ways to increase the distance over which quantum information can be reliably transferred and distributed. This is crucial for quantum cryptography, for quantum teleportation and for quantum networks,  key ingredients of quantum computers. The results presented in this Letter demonstrate that careful manipulation of the environmental properties based on the exploitation of environmental memory effects and non-Markovianity can be generally used to induce revivals of classical and quantum capacities as  well as for engineering distance-independent values of these quantities. This means that, in principle, non-Markovianity allows for realistic error correction schemes working for any channel length. In this sense non-Markovianity is a new and yet unexplored resource for quantum technologies, with the potential to pave the way to real-scale quantum-enhanced devices.

\section*{Acknowledgements}
S. M. acknowledges financial support from EPSRC (EP/J016349/1), the Finnish Cultural Foundation (Science Workshop on Entanglement), and the Emil Aaltonen foundation (Non-Markovian Quantum Information). B.B. thanks the Institute of Photonics and Quantum Sciences and the Open Quantum Systems and Entanglement group at Heriot-Watt University, Edinburgh, where part of this work was done, for the hospitality, and acknowledges financial support form the Emil Altonen Foundation and the National Science Center project 2011/01/N/ST2/00393. D. C. was partially supported by the National Science Center project DEC-2011/03/B/ST2/00136. The authors thank Shashank Virmani, Chiara Macchiavello and Pinja Haikka for valuable discussions on the topic of the Letter.

\section*{Supplementary Information}
In the Supplementary Information we expand on the three case studies presented in the Letter and further discuss the comparison between the newly introduced non-Markovianity measures $\mathcal{N}_Q$ and $\mathcal{N}_{C}$ and the ones previously introduced in the literature.

\subsection{Channels and capacities}

A quantum channel $\mathcal{E}$ is a completely positive and
trace preserving map  from an initial system state $\rho$ to the final state $\mathcal{E}(\rho)$ (Ref. \cite{Wilde} serves  as a self-contained modern text on quantum information theory). Any quantum channel may be modeled as a unitary evolution $U$ of the "system + environment" followed by the partial trace over an environmental degrees of freedom
\begin{equation}\label{}
    \mathcal{E}(\rho) = {\rm Tr}_{\rm E} ( U \, \rho \ot \omega\, U^\dagger) \ ,
\end{equation}
where $\omega$ denotes a fixed state of the environment. The above representation enables one to introduce a complementary channel
\begin{equation}\label{}
    \widetilde{\mathcal{E}}(\rho) = {\rm Tr}_{\rm S} ( U \, \rho \ot \omega\, U^\dagger) \ ,
\end{equation}
where now one performs a partial trace over the system degrees of freedom. Interestingly, taking the operator-sum representation
\begin{equation}\label{}
\mathcal{E}(\rho) = \sum_i K_i \rho K_i^\dagger\ ,
\end{equation}
of a channel $\mathcal{E}$ one finds the following simple representation of the complementary channel $\widetilde{\mathcal{E}}$
\begin{equation}\label{}
\widetilde{\mathcal{E}}(\rho) = \sum_{i,j} {\rm Tr}( K_i\, \rho \, K_j^\dagger) |i\>_E\<j|\ ,
\end{equation}
where   $|i\>_E$ is an orthonormal set in $\mathcal{H}_E$ (Hilbert space of the environment). A quantum channel is called degradable if it can be degraded to its complementary channel, that is, there exists another channel $\mathcal{F}$ such that $\widetilde{\mathcal{E}} = \mathcal{F}\, \mathcal{E}$.

In what follows we analyze two types of channel capacities: entangled-assisted classical capacity $C_{ea}$ and quantum capacity $Q$.
Recall that $C_{ea}$ is defined as the maximum amount of classical information reliably transmitted over the
quantum channel  if the sender and receiver share unlimited resources of entanglement.
This quantity is obtained by maximization of the quantum mutual information for single
channel use, which yields
\begin{equation}\label{Cea}
    C_{ea}(\mathcal{E}) = \max_\rho\Big[ S(\rho) +  S(\mathcal{E}(\rho)) - S([\oper \ot \mathcal{E}]|\psi\>\<\psi|) \Big] \ ,
\end{equation}
where $|\psi\> \in \mathcal{H}\ot \mathcal{H}_E$ is a purification of $\rho$, that is, $\rho = {\rm Tr}_{\rm E} |\psi\>\<\psi|$. Interestingly, the entropy of exchange $S(\rho,\mathcal{E}) \equiv S([\oper \ot \mathcal{E}]|\psi\>\<\psi|)$ does not depend on a particular purification and hence it is an intrinsic property of the pair $(\rho,\mathcal{E})$. If $\rho_*$ is a system state which maximizes formula (\ref{Cea}) then the entropy of entanglement $S(\rho_*)$ gives the amount of pure-state entanglement used by this entangled-assisted  protocol.

The quantum capacity $Q$ is the maximum amount of quantum
information transmitted by a quantum channel per channel
use. It is defined in terms of coherent information
\begin{equation}
    I_c(\rho,\mathcal{E}) = S(\mathcal{E}(\rho)) -  S(\rho,\mathcal{E}) \ .
\end{equation}
It can be proved that
\begin{equation}
    {I}_c(\rho,\mathcal{E}) = S(\mathcal{E}(\rho)) - S(\widetilde{\mathcal{E}}(\rho))\ .
\end{equation}
One defines quantum capacity of $\mathcal{E}$ as the following limit
\begin{equation}\label{}
    Q(\mathcal{E}) = \lim_{n\rightarrow \infty} \frac{Q_n}{n}\ ,
\end{equation}
where
\begin{equation}\label{}
    Q_n = \max_{\rho_n} \,  {I}_c(\rho_n,\mathcal{E}^{\ot n})\ ,
\end{equation}
and $\rho_n$ is a state in $\mathcal{H}^{\ot n}$ ($n$ copies of the original system Hilbert space $\mathcal{H}$).
It should be stressed that the limit $n\rightarrow \infty$  is necessary
as $I_c$ is super-additive, which makes the evaluation
of $Q$ difficult. However, for degradable channels the coherent
information is additive  and hence in this case $Q_n = n Q_1$ which shows that $Q = Q_1$.  Degradable channels provide, therefore, important class of channels  which allow in many cases  explicit calculation of quantum capacity.

\subsection{Dephasing Channel}

\paragraph{Dynamical map and its capacity}

The dephasing channel (dynamical map) $\Phi^D_t$ for a qubit is described by the
local in time master equation
\begin{equation}\label{}
    \frac{d}{dt} \Phi^D_t = L^D_t \Phi^D_t\ , \ \ \ \Phi^D_0=\oper\ ,
\end{equation}
with the following local generator
\begin{equation}\label{LD}
    L^D_t \rho=\frac{1}{2} \gamma(t) (\sigma_z \rho \sigma_z - \rho)\ ,
\end{equation}
and  time-dependent dephasing rate $\gamma(t)$.
This generator, and the corresponding master equation, can be derived exactly from the following microscopic Hamiltonian description of system (noisy channel) plus environment
 \begin{eqnarray}
 H= \omega_0 \sigma_z+ \sum_k  \omega_k a^{\dag}_k a_k + \sum_k  \sigma_z (g_k a_k+ g_k^* a^{\dag}_k), \nonumber
 \end{eqnarray}
with $\omega_0$ the qubit frequency, $\omega_k$ the frequencies of the reservoir modes, $a_k \;(a_k^{\dag})$ the annihilation (creation) operators of the bosonic environment and $g_k$ the coupling constant between each reservoir mode and the qubit. In the continuum limit $\sum_k |g_k|^2 \rightarrow \int d\omega J(\omega) \delta (\omega_k-\omega)$, where $J(\omega)$ is the reservoir spectral density \cite{Massimo,NMBEC}.

One easily finds the operator-sum representation $\Phi^D_t(\rho) = \sum_{i=1}^2 K_i(t) \rho K_i^\dagger(t)$ with the time-dependent Kraus operators: $K_1(t)=\sqrt{\frac{1+e^{-\Gamma(t)}}{2}}\, \mathbb{I}$ and $K_2(t)=\sqrt{\frac{1-e^{-\Gamma(t)}}{2}}\, \sigma_z$, where $\Gamma(t)=\int_0^t \gamma(\tau) d\tau$.
While $\gamma(t)$ may temporarily attain negative values, complete positivity of the dynamical map imposes that $\Gamma(t) \ge 0$. We note in passing that the usual Markovian quantum channel for pure dephasing can be written in terms of Kraus operators having the same operatorial form as those reported above, provided one replaces $\Gamma(t)$ with $\gamma_M t$.

Using the Kraus operators one can write the complementary map, needed to calculate both the coherent information and the entropy exchange which appears in the definition of the mutual information of the channel, as follows:
\begin{eqnarray*}
  \widetilde{\Phi_t^D}(\rho) &=& \frac 12 \Big[ (1+e^{-\Gamma(t)}) |1\>_{\rm E}\<1| + (1-e^{-\Gamma(t)}) |2\>_{\rm E}\<2| \Big]\nonumber  \\
  &+&  \frac 12  \sqrt{1-e^{-2\Gamma(t)}}\, {\rm Tr}(\rho \sigma_z)\,  \Big( |1\>_{\rm E}\<2| + |2\>_{\rm E}\<1| \Big) \ .
\end{eqnarray*}
The dephasing channel is degradable for all values of
$\Gamma(t)$, which simplifies the calculations of the quantum
capacity. In this case, indeed, we find that the state
optimizing the coherent information in the definition of
the quantum capacity does not depend either on time
or on the specific properties of the environmental spectrum. Having this in mind one can show that $Q$ takes
the following simple analytical formula \cite{Wilde}
\begin{equation}\label{}
    Q^D(t) = 1- H_2\left( \frac{1 + e^{-2\Gamma(t)}}{2} \right) ,
\end{equation}
with $H_2(\, . \, )$ the binary Shannon entropy. Since
\begin{equation}\label{}
    \frac{d}{dt} Q^D(t) = -\frac 12\, \gamma(t) \,e^{-\Gamma(t)}  \log_2 \frac{1 + e^{-2\Gamma(t)}}{1 - e^{-2\Gamma(t)}}
\end{equation}
the measure $\mathcal{N}_Q$ has nonzero value if and only if $\gamma(t) < 0$, i.e., whenever the dynamical map $\Phi_t$ is not divisible.


\paragraph{Comparison between Non-Markovianity Measures}

As mentioned in the Letter, for the pure dephasing channel all known non-Markovianity measures consistently detect the Markovian - non-Markovian crossover and are therefore equivalent to divisibility. Stated another way, in this case, divisibility is a necessary and sufficient condition for all forms of non-Markovianity. The two measures we introduce in this Letter, however, present some advantages with respect to, e.g., the BLP measure introduced in Ref. \cite{NMBLP}. The first advantage is not specific to this model but general. Indeed, both ${\cal N}_Q$ and ${\cal N}_{C}$ are based on the definition of the capacities $Q$ and $C_{ea}$, respectively, which in turn are obtained by optimizing over the input state $\rho$ only, while the BLP measure requires optimization over pairs of initial states, making it much more complicated to calculate, even numerically. Increasing the number of qubits makes the computation of such measure practically intractable. The second advantage is specific to the pure dephasing model. As this is a degradable channel, additivity of both $Q$ and $C_{ea}$ holds. This allows to reduce the calculation of ${\cal N}_Q$ and ${\cal N}_{C}$ for $n$ qubits subjected to independent local dephasing to the one qubit analytical formula. We note that this is not true for all the other non-Markovianity measures which, generally, are neither additive nor subadditive. As a consequence, due to the increasing complexity in solving the optimization problem for increasing numbers of qubits, both the BLP and the RHP  \cite{NMRHP}  measures have been until now calculated (numerically) for up to 2 qubits.

\subsection{Amplitude Damping Channel}

\paragraph{Dynamical map and capacities}

The amplitude damping channel $\Phi^A_t$ is described by the following exact local generator:
\begin{equation}\label{LAD}
L^A_t \rho=-\frac{i s(t)}{2} [\sigma_+ \sigma_-, \rho]+\gamma(t) \Big(\sigma_- \rho\sigma_+ -\frac{1}{2}\{\sigma_+ \sigma_-,\rho\} \Big),
\end{equation}
with $s(t)=-2 \Im\{\dot{G}(t)/G(t)\}$ and $\gamma(t)=-2 \Re\{\dot{G}(t)/G(t)\}$ the time-dependent Lamb shift and decay rate, respectively, where $G(t)$ satisfies the non-local equation $\dot{G}(t)=- \int_0^t f(t-t') G(t') dt'$ with initial condition $G(0)=1$, and $f(t)$ is the reservoir correlation function which is related {\em via} the Fourier transform with a spectral density $J(\omega)$. As usual $\sigma_+$ and $\sigma_-$ are standard qubit raising and lowering operators, respectively.
The local generator and corresponding master equation can be derived exactly by the following microscopic Hamiltonian model describing a two-state system interacting with a bosonic quantum reservoir at zero temperature \cite{BPBook}
\begin{equation}\label{Hamiltonian}
H=\omega_{0} \sigma_z+\sum_{k} \omega_{k}a^{\dagger}_{k}a_{k}+\sum_{k}(
g_{k}a_{k}\sigma_{+}+g_{k}^*a_{k}^{\dag}\sigma_{-}).
\end{equation}
The evolution is described by the following density operator:
\begin{equation}
    \Phi^A_t(\rho)=\left(
              \begin{array}{cc}
                1-|G(t)|^2 \rho_{22} & G(t) \rho_{12} \\
                G^*(t) \rho_{12}^* & |G(t)|^2 \rho_{22} \\
              \end{array}
            \right).
\end{equation}

The generator of the Markovian amplitude damping channel has the same form of Eq.(\ref{LAD}), provided that one replaces the time dependent coefficients $s(t)$ and $\gamma(t)$ with positive constant values $s_M$ and $\gamma_M$.

The Kraus representation $\Phi^A_t(\rho) = \sum_{i=1}^2 K_i(t) \rho K_i^\dag(t)$ for the amplitude damping channel is given by
$K_1=\left(
       \begin{array}{cc}
         1 & 0 \\
         0 & G(t) \\
       \end{array}
     \right)
$
and
$K_2=\left(
       \begin{array}{cc}
         0 & \sqrt{1-|G(t)|^2} \\
         0 & 0 \\
       \end{array}
     \right)
$
which gives us a complementary map defined by:
\begin{eqnarray*}
   \widetilde{\Phi^A_t}(\rho) &=& [1 - (1- |G(t)|^2) \rho_{22} ] |1\>_{\rm E}\<1| \nonumber \\ &+& (1-|G(t)|^2)\,\rho_{22} |2\>_{\rm E}\<2| \\
  &+&   \sqrt{1- |G(t)|^2}\, \big( \rho_{12}|1\>_{\rm E}\<2| + \rho_{21}|2\>_{\rm E}\<1| \big)
   \ .
\end{eqnarray*}
One finds \cite{Wilde} the following formulae for classical entangled-assisted capacity $C_{ea}^A := C_{ea}(\Phi^A_t)$
\begin{equation*}\label{}
    C_{ea}^A  = \max_{p\in [0,1]} \Big\{ H_2(p) + H_2(|G(t)|^2p) -   H_2([1-|G(t)|^2]p)\Big\} \ ,
\end{equation*}
and quantum capacity $Q^A := Q(\Phi^A_t)$
\begin{equation*}\label{}
    Q^A = \max_{p\in [0,1]} \Big\{ H_2(|G(t)|^2p) -   H_2([1-|G(t)|^2]p)\Big\} \ ,
\end{equation*}
for $|G(t)|^2 >  \frac 12$ (otherwise $Q(\Phi^A_t)\equiv 0$). In the above formulae one still needs to perform a simple optimization over probability $p\in [0,1]$.

\paragraph{Lorentzian Spectrum}

If the reservoir spectral density has a Lorenztian shape, i.e. $J(\omega)=\gamma_M \lambda^2/2 \pi [(\omega-\omega_c)^2+\lambda^2]$, then the function $G(t)$ takes the form
\begin{eqnarray}
 && G(t) = e^{-\frac{(\lambda-i\delta)t}{2}}
\left [\cosh \left(\frac{\Omega t}{2} \right) + \frac{\lambda-i
\delta}{\Omega} \sinh \left(\frac{\Omega t}{2} \right)
\right ]\, \label{eq:e},
\end{eqnarray}
with
$$\Omega= \sqrt{\lambda^2 - 2 i \delta
\lambda- 4 w^2},$$ where
$w=\gamma_M \lambda / 2+\delta^2/4$, and $\delta = \omega_0 - \omega_c$.

For $\delta = 0$, one obtains the following solution
\begin{eqnarray}
    G(t)&=& e^{-\lambda t/2} \left[\cosh \left(\sqrt{1-2R} \frac{\lambda t}{2} \right) \right. \nonumber\\
    &+& \left.\frac{1}{\sqrt{1-2R}} \sinh \left(\sqrt{1-2R} \frac{\lambda t}{2} \right)\right],
 \end{eqnarray}
 with $R=\gamma_M/\lambda$.
In the weak coupling regime, i.e. for $R\le \frac{1}{2}$, $G(t)$ is monotonically decreasing, whereas in the strong coupling regime, $R > \frac{1}{2}$, $G(t)$ is oscillating so there are periods when $\frac{d}{dt}|G(t)|$ is positive. It is straightforward to show that $\gamma(t) \ge 0$ for $R\le \frac{1}{2}$, while it can take temporarily negative values for $R > \frac{1}{2}$. In the latter case the dynamical map is not divisible.
It is well known that the amplitude damping channel is degradable only for $|G(t)|^2 >\frac{1}{2}$, otherwise it is a non-degradable channel with zero quantum capacity. Hence, from a quantum information processing point of view, revivals of $|G(t)|^2$ are important only when they occur in the region $|G(t)|^2>\frac{1}{2}$.

\paragraph{Photonic Band Gap}
Also the second example of amplitude damping channel stems, as the previous one, from an exact microscopic model of an open quantum system. The environment is a bosonic zero temperature three-dimensional periodic dielectric with isotropic photon dispersion relation. In this ideal photonic crystals, a photonic band gap is the frequency range over which the local density of electromagnetic states and the decay rate of the atomic population of the excited state vanish. Near the band gap edges the density of states becomes singular, the atom-field interaction becomes strong, and one can expect modifications to the spontaneous emission decay. We consider the model described in Ref. \cite{PBG}. In this case the function $|G(t)|^2$ depends on two relevant parameters, the detuning $\delta=\omega_0 - \omega_e$ from the band gap edge frequency $\omega_e$ and the parameter $\beta $ defined as $\beta^{3/2} = \omega_0^{7/2} d^2/ 6 \pi \epsilon_0 \hbar c^3$ with $\epsilon_0$ the Coulomb constant and $d$ the atomic dipole moment. Population trapping is a general feature of this model. The farther the atomic transition frequency is from $\omega_e$ and inside the gap, the higher is the fraction of initial state population which is trapped in the two-state system for $t\rightarrow \infty$. This in turn gives rise to stationary values of the quantum and classical capacities as we have shown in the Letter.

\paragraph{Comparison between Non-Markovianity Measures}

\begin{figure}
\includegraphics[width=0.45\textwidth]{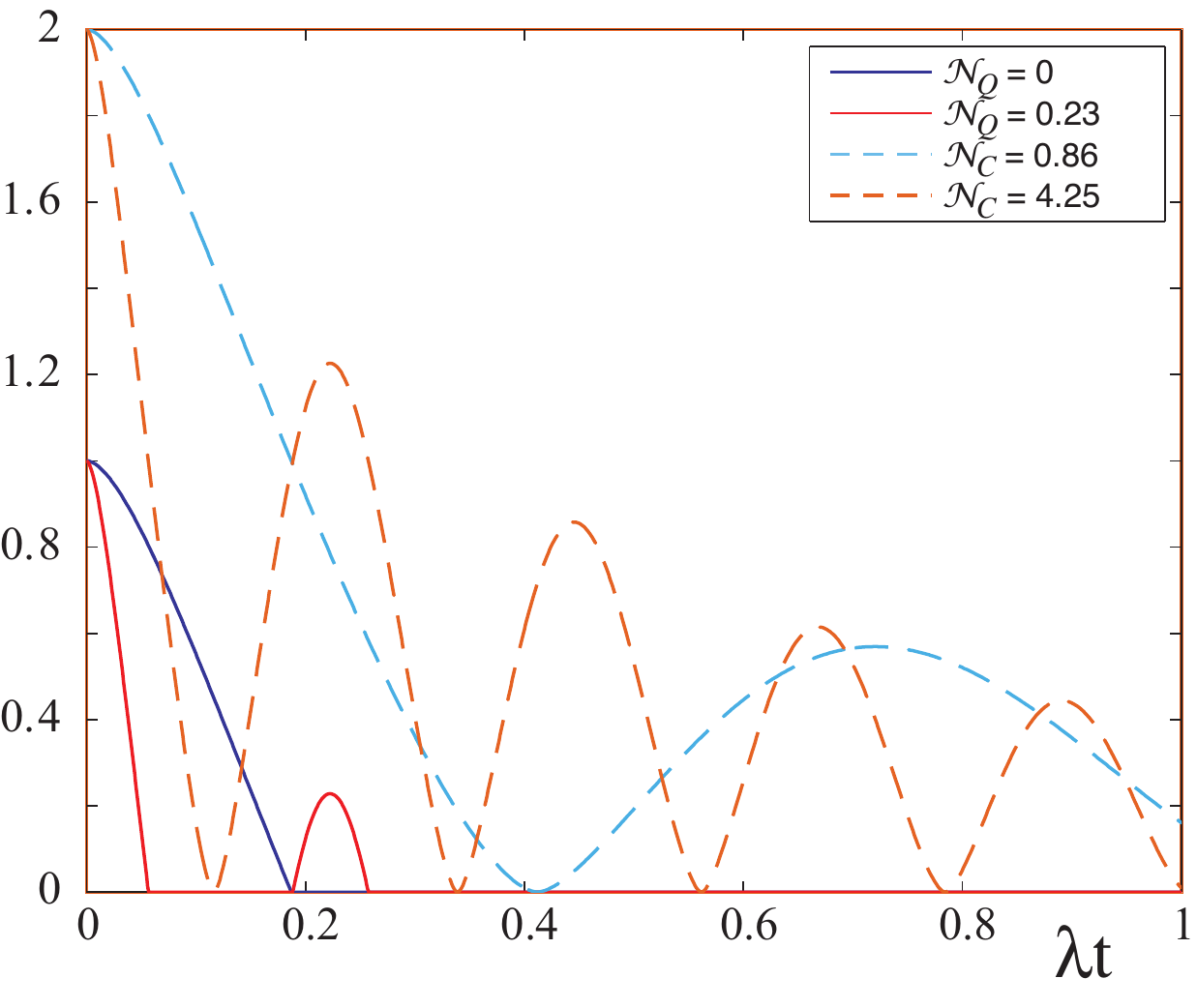}
\caption{{Quantum channel capacity $Q^A$ and entanglement assisted capacity $C_{ea}$ as a function of time or, equivalently, of the channel length for the exact amplitude damping model with Lorentzian reservoir spectrum}. In the plot we have set $\delta=0$, and compared the capacities for $R = 10$  and $R=100$. Values of the non-Markovianity measures for the two cases are reported in the inset.}
\end{figure}

\begin{figure}
\includegraphics[width=0.45\textwidth]{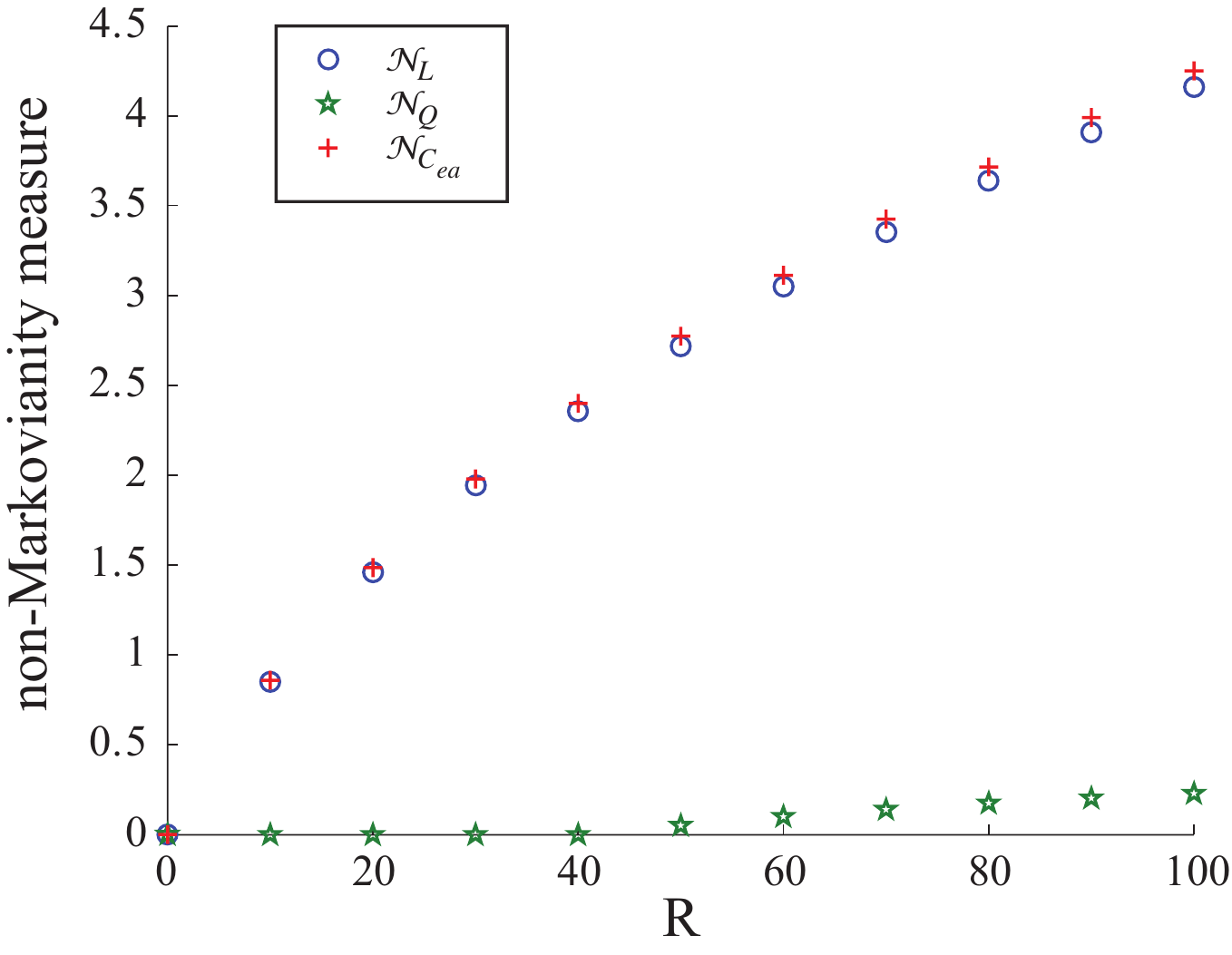}
\caption{Non-Markovianity measures $\mathcal{N}_L$ and  $\mathcal{N}_{C}$ for the amplitude damping channel with Lorentzian spectrum on resonance $\delta=0$, for increasing values of $R$  in the strong coupling regime.}
\end{figure}

In Fig. 1 we plot the entanglement assisted capacity $C^A_{ea}$ and the quantum capacity $Q^A$ for the amplitude damping channel with Lorentzian spectrum and for two different values of $R$ in the strong coupling regime, i.e. for $R = 10$ and $R=100$. For these values of parameters the dynamical map is always non-divisible and, hence, all non-Markovianity measures introduced in Ref. \cite{NMBLP} (BLP), Ref. \cite{NMRHP} (RHP), Ref. \cite{NMWolf} (WCEC), and Ref. \cite{NMLuo} (LFS) are non-zero. However, in Fig. 1 we see that, while $C_{ea}$ always exhibits revivals, i.e. ${\cal N}_{C} \neq 0$, $Q^A$ decreases monotonically, eventually vanishing for $R=10$, hence in this case ${\cal N}_{Q} = 0$. Revivals of ${\cal N}_{Q}$ occurs only for greater values of $R$ ($R=100$, in Fig.1). This example shows the difference between different non-Markovianity measures, and in particular between the two introduced in this Letter. The careful reader will not be surprised by this result. It is expected that the transmission of quantum information along a quantum channel is more sensitive to noise than the transmission of classical information (although assisted by entanglement shared between Alice and Bob). Hence, revivals of $Q$ due to system-environment memory  might require a stronger condition than revivals of $C_{ea}$, in the case of the example here considered a higher value of the system-reservoir coupling constant. Once more, this is consistent with our general viewpoint: different measures quantify different properties, all stemming from system-reservoir memory, which are useful for different tasks or for different protocols.

In Fig. 2 we make a more explicit comparison between the two measures introduced in this Letter and  the LSF measure \cite{NMLuo} as it is the one which appears to be more closely connected to ${\cal N}_{C}$. This measure stems from a property of the quantum mutual information, namely the fact that for divisible maps $I((\Phi_t \ot \mathbb{I}) \rho^{AB}) \le I((\Phi_s \ot \mathbb{I}) \rho^{AB})$, for any time $0 \le s \le t$, and for any bipartite state $\rho^{AB}$. This property follows from the monotonicity of the mutual information under local operations.

The LSF non-Markovianity measure detects a partial and temporary increase in the correlations between a system undergoing a non-unitary evolution and an ancilla, as measured by mutual information, and it is defined as follows

\begin{equation}\label{NL}
    \mathcal{N}_L(\Phi_t)=\sup_ {\rho^{SA}}\int_{\frac{d}{dt} I(\rho_t^{SA})>0} \frac{d}{dt} I(\rho_t^{SA}) dt,
\end{equation}
where $\rho^{SA}$ is the initial system-ancilla state and $\rho^{SA}_t = ({I} \otimes \Phi_t) \rho^{SA}$.
Here the optimization is done over all possible initial states of system and ancilla.



It is worth noticing that, although both ${\cal N}_{C}$ and $ \mathcal{N}_L$ are defined in terms of the mutual information, and $ I(\rho_t^{SA})$ coincides with $I(\rho,\Phi_t)$ used in the definition of $C_{ea}$, the maximization over the initial state is not the same. More precisely, the optimal states of $C_{ea}$ do not coincide with the states optimizing the integral in Eq. (\ref{NL}). In Fig. 2 we compare the LFS non-Markovianity measure $ \mathcal{N}_L$ to ${\cal N}_{C}$ and $ \mathcal{N}_Q$ for the amplitude damping channel with Lorentzian spectral density on resonance ($\delta=0$).
One has $ \mathcal{N}_Q =0$ for $R \lesssim 43$.

\end{document}